\documentclass[11pt]{article}

\usepackage[T1]{fontenc}
\usepackage{mathptmx}                    
\usepackage[a4paper, margin=1in]{geometry}
\usepackage{setspace}

\usepackage{amsmath}
\usepackage{amssymb}
\usepackage{amsthm}

\usepackage{bm}

\usepackage{graphicx}
\usepackage{float}
\usepackage{booktabs}
\usepackage{array}
\usepackage{longtable}
\usepackage{multirow}
\usepackage[most]{tcolorbox}

\usepackage[round,authoryear]{natbib}

\usepackage{xcolor}
\definecolor{brandburgundy}{RGB}{128,0,32}

\usepackage{url}
\usepackage{hyperref}
\newcommand{\doi}[1]{\href{https://doi.org/#1}{\nolinkurl{#1}}}
\hypersetup{
    colorlinks=true,
    linkcolor=brandburgundy,
    citecolor=brandburgundy,
    urlcolor=brandburgundy,
    breaklinks=true,
    pdftitle={Are Whitepaper Claims Reflected in Market Structure? A Contamination-Aware Pipeline and a Power-Limited Null},
    pdfauthor={Murad Farzulla},
    pdfkeywords={cryptocurrency, NLP, narrative economics, power analysis, corpus contamination}
}

\def\UrlBreaks{\do\/\do-\do_}
\expandafter\def\expandafter\UrlBreaks\expandafter{\UrlBreaks%
  \do\a\do\b\do\c\do\d\do\e\do\f\do\g\do\h\do\i\do\j\do\k%
  \do\l\do\m\do\n\do\o\do\p\do\q\do\r\do\s\do\t\do\u\do\v%
  \do\w\do\x\do\y\do\z}

\usepackage{enumitem}
\usepackage{fancyhdr}

\usepackage{titlesec}
\titleformat{\section}{\normalfont\large\bfseries\color{brandburgundy}}{\thesection}{0.5em}{}
\titleformat{\subsection}{\normalfont\normalsize\bfseries\color{brandburgundy}}{\thesubsection}{0.5em}{}
\titleformat{\subsubsection}{\normalfont\small\bfseries\color{brandburgundy}}{\thesubsubsection}{0.5em}{}
\titlespacing*{\section}{0pt}{2ex plus 0.8ex minus 0.2ex}{1ex plus 0.3ex}
\titlespacing*{\subsection}{0pt}{1.5ex plus 0.5ex minus 0.2ex}{0.8ex plus 0.2ex}
\titlespacing*{\subsubsection}{0pt}{1.2ex plus 0.4ex minus 0.2ex}{0.6ex plus 0.2ex}

\fancypagestyle{firstpage}{
  \fancyhf{}
  \fancyfoot[C]{\small\thepage}

}

\theoremstyle{definition}

\theoremstyle{plain}

\newcommand{\papernum}{DAI-2508}              

\newcommand{\paperdate}{12 September 2026}

\newcommand{\paperurl}{https://systems.ac/3/DAI-2508}  

\begin{document}

\thispagestyle{firstpage}

\begin{center}
{\small\textsc{\href{https://dissensus.ai}{Dissensus} Working Paper Series}}\\[0.2em]
{\small \href{\paperurl}{\papernum}}
\end{center}

\vspace{1.5em}

\begin{center}
{\LARGE\bfseries Are Whitepaper Claims Reflected in Market Structure?}\\[0.5em]
{\large\itshape A Contamination-Aware Pipeline and a Power-Limited Null}\\[1.5em]

{\large Murad Farzulla}\textsuperscript{1,*}\\[0.8em]

{\small
  \textsuperscript{1}\href{https://dissensus.ai}{Dissensus}, London, UK%
}\\[0.5em]

{\footnotesize
  \textsuperscript{*}Correspondence: \href{mailto:murad@dissensus.ai}{murad@dissensus.ai}
  \quad
  ORCID: \href{https://orcid.org/0009-0002-7164-8704}{0009-0002-7164-8704}%
}\\[0.3em]
{\footnotesize \paperdate}
\end{center}

\begin{abstract}
Do the functional narratives in cryptocurrency whitepapers correspond to their tokens' market behaviour? We compare ten-category topical-emphasis profiles for 43 screened documents with seven market statistics calculated from 2023--2024 exchange data. The primary specification reconstructs USD notional turnover from hourly bars. Dimension-matched Procrustes congruence is $\phi=0.280$ (permutation $p=0.507$), slightly below its permutation-null mean of $0.284$; the zero-padded statistic gives the same non-detection. A four-leg comparison separates document replacement from changes in the assets included. Replacing documents on the 34 common assets changes padded congruence by $-0.014$ under USD turnover and $-0.009$ under base-token volume. Entity rankings and threshold crossings depend on both composition and specification, so an earlier contamination-only attribution is withdrawn. The documents are not a verified historical corpus: at least two postdate the market window. Excluding these documents, or excluding all seven assets with shorter histories, does not produce a significant alignment. Fresh numerical simulations distinguish injected signal from fitted congruence and compare noise restricted to the market subspace with noise throughout the text space. At the lowest classifier-agreement scenario, detection remains below $43\%$ even at the largest injected signal. These are conditional checks of the alignment stage, not validation of the text instrument or exclusion bounds on economic effects. The contribution is an auditable non-detection and a specification-sensitive corpus diagnosis, with the inferential limits made explicit.


\end{abstract}
\setstretch{1.15}

\section{Introduction}
\label{sec:intro}
Cryptocurrency whitepapers describe projects' intended functions, technical architectures and use cases. They are therefore an attractive source for studying how project narratives relate to market behaviour. Their availability does not, however, make either the documents or a machine-derived representation of their content a validated measure of economic commitments. A downloaded file can describe the wrong project; a genuine document can postdate the period being studied; and a classifier can register discussion of a topic without identifying an assertion about that topic.

We ask a restricted empirical question: does a cross-sectional representation of document topics align more closely with a cross-sectional representation of token-market behaviour than would arise from randomly pairing their rows? This question differs from whether narratives predict returns, whether markets efficiently incorporate information, or whether projects fulfil their promises. The narrative-economics literature motivates attention to textual information \citep{shiller2017narrative}; studies of cryptocurrency returns identify crypto-specific time-series momentum and investor-attention patterns \citep{liu2021risks} and common market, size and momentum factors \citep{liu2022common}. None establishes that the particular functional categories used here should map onto the seven market statistics in our analysis.

The study combines natural-language-inference classification of 43 project documents with daily summaries constructed from hourly market data for 2023--2024. Orthogonal Procrustes rotation aligns the resulting spaces. We evaluate a centred, correlation-type congruence statistic against a permutation null that repeats the fitting operation. With USD turnover as the primary volume measure, the dimension-matched statistic is $0.280$ and its permutation-null mean is $0.284$. The test does not detect alignment. The difference between these values also shows why a positive fitted coefficient cannot be read as evidence of a small substantive effect: fitting a rotation to unrelated finite samples already produces positive congruence.

A second contribution is a correction to an earlier interpretation of this dataset. The original corpus snapshot contained eleven problematic documents among 37 analysed assets. Earlier drafts attributed an apparent entity ordering entirely to contamination. Recomputing the comparison while holding the asset set fixed shows that replacing documents, dropping assets and adding assets have distinct effects. The numerical effects and entity threshold crossings also depend on the volume specification. We therefore replace the singular causal story with a reproducible diagnosis of sensitivity to corpus composition, document content and market measurement. This is one dataset-specific finding, not evidence that one failure mode generally dominates the others.

Finally, we test the numerical alignment stage with simulations on the primary USD-statistics matrix. The simulations expose the assumptions required for interpreting detection rates. Their input signal parameter is not the fitted coefficient, their classifier-agreement parameter is assumed rather than estimated against ground truth, and their noise model matters. They establish conditional sensitivity of a numerical procedure. They do not validate the text classifier, identify an economic effect, or exclude a population alignment from a non-significant result.

\section{Related Work and Scope}
Zero-shot classification can formulate a labelling problem as natural-language entailment \citep{yin2019benchmarking}. BART supplies a sequence-to-sequence pretrained model \citep{lewis2020bart}, while the MNLI benchmark supplies natural-language-inference supervision \citep{williams2018broad}. These are distinct contributions: a model's ability to perform entailment classification does not establish that a researcher-selected label set captures a finance construct. We therefore distinguish the computational definition of our scores from the validity of the construct they are intended to approximate.

Our financial representation is also selective. Cross-sectional cryptocurrency factor research provides a reason to consider return, risk and trading-activity variation \citep{liu2022common}, but it does not derive our seven-statistic representation or predict a rotation linking it to whitepaper topics. We analyse a descriptive representation of market behaviour. A failure to align with this representation does not establish a failure to align with liquidity measures, network adoption, protocol activity, event-time returns, derivatives markets or other omitted outcomes.

The matrix comparison draws on the orthogonal Procrustes solution of \citet{schonemann1966generalized} and on congruence comparisons discussed by \citet{lorenzo2006tuckers}. Thresholds developed for comparing factor solutions from similar data do not automatically transfer to heterogeneous text and market spaces. We use the fitted statistic's own random-pairing distribution for inference and assign no universal ``weak'', ``moderate'' or ``strong'' interpretation to a numerical coefficient. The contribution lies in the application and its auditable limits, rather than a new alignment estimator.

\section{Data and Corpus Provenance}
\label{sec:data}
The market archive contains hourly OHLCV observations for 49 assets collected through CCXT from Binance. Its broad observation window runs from 1 January 2023 to 31 December 2024; the longest archived histories contain 17,543 rows. The aligned sample comprises 43 assets with stored ten-category text profiles. Selection depends on exchange coverage, liquidity and document availability, and is not a random sample of cryptocurrency projects. The 43-asset list and each history's date range appear in the release evidence.

Seven aligned assets have fewer observations than the longest histories: ARB, OCEAN, POL, RENDER, RPL, SUI and XMR. Statistics for these assets use their available histories, which differ both in length and in market regime. The headline sample therefore does not provide a uniform two-year history for every token. We report a sensitivity analysis excluding all seven assets. Cross-sectional means and standard deviations used for market scaling are retained from the 49-asset universe in these exclusions, so the perturbation changes the analysed rows rather than silently redefining the measurement scale.

\subsection{Document identity and historical availability}
The screened corpus contains whitepapers, protocol specifications and technical documentation. The earlier 37-asset snapshot included eleven files subsequently replaced or excluded: eight assets have replacement text on the same asset set (ADA, ALGO, ATOM, BAND, DOT, GRT, NEAR and RPL), while AXS, SUSHI and YFI have no counterpart in the screened sample. The screened sample additionally contains nine assets absent from the earlier sample: ENJ, FTM, IMX, LPT, OP, POL, RENDER, SNX and TRB. Its larger size consequently reflects collection and replacement as well as exclusion.

The word ``screened'' describes content checks and repairs in that corpus-building process; it does not certify a complete historical provenance chain. The legacy metadata file retains some pre-replacement word counts and flags, despite pointing to newer files. The release therefore supplies a separate manifest linking each analysed symbol to the current document path, SHA-256 hash, stored extracted chunks and chunk hash. Historical metadata flags are retained explicitly as historical fields, rather than treated as current verification verdicts. The manifest makes the frozen inputs inspectable; file hashes do not by themselves prove that a document is authentic, complete or historically available.

At least two documents postdate the market window. The stored OP text begins with an April 2025 MiCA whitepaper prepared and filed by LCX, and the ZEC protocol specification identifies a January 2026 version. These are not 2023--2024 founding narratives. Other documents span different vintages, and a systematic version-date verification has not been completed for every asset. We therefore describe the design as a cross-sectional comparison using heterogeneous document vintages, not as a historical forecasting study or a fully contemporaneous narrative test. Excluding OP and ZEC is a sensitivity check for two known mismatches, not a guarantee that the remaining 41 documents are temporally valid.

\section{Methods}
\label{sec:methods}
\subsection{Topical-emphasis profiles}
The primary text representation is the archived output of \texttt{facebook/bart-large-mnli}. The analysis uses ten author-selected categories: store of value, medium of exchange, smart contracts, decentralised finance, governance, scalability, privacy, interoperability, data storage and oracle services. These are exploratory functional categories, not a taxonomy adopted from a validated financial measurement scale.

The extractor batches sentences by character count, targeting chunks between 100 and 1,000 characters. It does not forcibly split an individual overlong sentence, so some chunks exceed the nominal upper threshold. This corrects earlier descriptions of 500-word chunks. The classifier passes each chunk and the ten complete label sentences from \texttt{taxonomy.py} to the Transformers zero-shot pipeline with \texttt{multi\_label=True}; for example, one label is ``This text describes privacy or anonymous transactions.'' No custom hypothesis template is passed. The pipeline's model-dependent tokenisation and truncation remain relevant for overlong inputs.

Let $s_{atk}$ denote the non-negative score assigned to category $k$ for chunk $t$ of asset $a$. The stored document profile is
\begin{equation}
 c_{ak}=\frac{|T_a|^{-1}\sum_{t\in T_a}s_{atk}}
 {\sum_{j=1}^{10}|T_a|^{-1}\sum_{t\in T_a}s_{atj}},
 \qquad \sum_k c_{ak}=1.
\end{equation}
Each chunk receives equal weight in this aggregation. The closing operation removes overall score magnitude and induces dependence between categories. A short or problematic document still yields a nonzero composition; it is not a near-zero vector merely because its source text is sparse. We do not interpret a high score as a verified project promise, a positive assertion, or a calibrated probability that a function is implemented. The measured construct is relative topical emphasis. There is no established upper- or lower-bound relationship between its alignment and the alignment of actual economic commitments.

The September verification uses the frozen scores and documents; it does not re-execute the large language models. This separation permits an exact check of the numerical analysis without conflating it with model-version drift. Full text-to-score replication requires the model stack and its weights; the release documents this additional dependency and preserves the analysed scores.

\subsection{Market statistics and units}
For each asset we aggregate hourly bars into daily closing prices and daily notional turnover,
\begin{equation}
 V_{ad}=\sum_{h\in d}v_{ah}P_{ah},\qquad
 r_{ad}=\log(P_{ad}/P_{a,d-1}).
\end{equation}
Here $v_{ah}$ is exchange volume in base-token units and $P_{ah}$ is the hourly close. The archive uses stablecoin quote pairs; USD notional is an approximation based on treating the quote unit as a dollar, not a trade-level USD volume record. We compute annualised mean log return, annualised return standard deviation, Sharpe ratio, absolute maximum drawdown, $\log(1+\overline V_a)$, the standard deviation of $V_{ad}$ divided by its mean, and a price-on-time slope divided by mean price. The trend statistic is consequently invariant to a change of price units. Returns and volatility use the 252-day convention. Changing that common positive annualisation constant cancels after each feature is standardised cross-sectionally.

Base-token volume cannot be compared across assets as though a BTC and a LINK unit were equal economic quantities. Replacing base volume with notional turnover moves Bitcoin from rank 47 to rank 1 among the 49 assets and Ethereum from rank 39 to rank 2. This is why the notional specification is primary. We also report the legacy base-token specification and the approximation using $(H+L+C)/3$ in place of close. The latter is a typical-price check, not an observed volume-weighted average price.

The seven columns are z-scored over the 49-asset market universe before selecting the aligned sample. In the 43-asset primary sample their covariance eigenvalues yield participation-ratio effective rank $3.07$, where $r_{\rm eff}=(\sum_j\lambda_j)^2/\sum_j\lambda_j^2$. Thus seven columns do not represent seven independent economic dimensions. The permutation null retains their covariance, but this does not expand the substantive scope of the chosen statistics. We do not treat turnover as observed market capitalisation. Regressing the market matrix on one of its own volume columns would also remove that column mechanically and would not constitute an independent size control.

\subsection{Alignment estimands and permutation inference}
For the primary statistic, principal components reduce the ten-column topic matrix to seven dimensions. Both this matrix $A$ and the seven-column market matrix $B$ are column-centred. We solve
\begin{equation}
 Q^*=\arg\min_{Q^\top Q=I}\|AQ-B\|_F^2,
 \qquad A^\top B=U\Sigma V^\top,\quad Q^*=UV^\top,
\end{equation}
and average the signed column correlations between $AQ^*$ and $B$:
\begin{equation}
 \phi_{\rm matched}=\frac{1}{7}\sum_{j=1}^{7}
 \frac{(AQ^*)_j^\top B_j}{\|(AQ^*)_j\|\,\|B_j\|}.
\end{equation}
The PCA projection selects high-variance text directions; low-variance topic information can be discarded. The rotation minimises a Frobenius-distance objective, not the mean column-correlation objective subsequently reported. Its direction is fixed: text is the source and market statistics are the target. These choices define the tested statistic and are repeated under permutation; they are not claims of unique or maximally attainable cross-domain alignment.

For continuity we also compute the legacy statistic: pad the seven-column market target with three zero columns, rotate the unreduced ten-column text matrix, and average absolute column congruences over all ten columns, assigning zero to padded target columns. This quantity is exactly $7/10$ times the same rotation's mean over informative target columns. The rescaling affects neither the permutation ordering nor its $p$-value. It is not a lower bound on economic alignment and is not comparable directly with the separately fitted primary statistic. Both are centred congruences rather than the classical uncentred Tucker coefficient.

We draw 1,000 row permutations of the market matrix using a local random generator with seed 42, refitting the rotation each time. The primary test compares absolute signed statistics; the legacy test uses the upper tail of the non-negative padded statistic. In this release $p=(b+1)/(B+1)$, where $b$ is the number of permutation statistics at least as extreme as observed and $B=1000$. The evidence also records the previous $b/B$ convention so historical numbers can be reconciled. Differences are at most 0.001 and change no conclusion. The null is exchangeability of text--market pairings conditional on the observed matrices. It is not a sector-adjusted or causal null; sector dependence, selection and omitted covariates limit a population interpretation.

We report the permutation mean alongside the observation because in-sample rotation fitting creates a positive chance level. Percentile bootstraps are omitted as inferential bounds: row duplication and refitting shift their distribution upward. The archived bootstrap is a diagnostic of this behaviour, not a confidence interval establishing an excluded effect size.

\section{Results}
\label{sec:results}
\subsection{The primary comparison does not exceed chance alignment}
Table~\ref{tab:primary} reports both estimands and their own null means. The primary statistic is $0.280$, against a null mean of $0.284$, giving null-centred excess $-0.0035$. The padded statistic is $0.210$, against $0.214$, giving excess $-0.0045$. These are non-detections. They do not estimate a small positive structural relation just because the fitted coefficients are positive.

\begin{table}[H]\centering
\caption{Primary USD-turnover comparison, $n=43$.}\label{tab:primary}
\begin{tabular}{lrrrr}\toprule
Statistic & Observed & Null mean & Null median & $p$\\\midrule
Dimension-matched & 0.280 & 0.284 & 0.281 & 0.507 \\
Zero-padded & 0.210 & 0.214 & 0.213 & 0.546 \\
\bottomrule\end{tabular}
\end{table}

Under base-token volume the matched coefficient is $0.303$ and the padded coefficient is $0.223$. The choice of economic units therefore changes the point estimates but does not reveal a significant relation under either specification. Using typical price to approximate hourly turnover gives matched $\phi=0.2803$ and $p=0.5065$ before rounding. The main conclusion is also unchanged when Bitcoin, the two known postperiod documents, or all shorter histories are excluded (Table~\ref{tab:sensitivity}). These are finite-sample checks, not proof that the remaining selection and vintage limitations are harmless.

\begin{table}[H]\centering
\caption{Release sensitivity checks on the primary numerical analysis.}\label{tab:sensitivity}
\begin{tabular}{lrrrr}\toprule
Variant & $n$ & Matched $\phi$ & Null mean & $p$\\\midrule
USD close, primary & 43 & 0.280 & 0.284 & 0.507 \\
USD typical price & 43 & 0.280 & 0.284 & 0.506 \\
Legacy base-token units & 43 & 0.303 & 0.290 & 0.366 \\
Exclude Bitcoin & 42 & 0.277 & 0.289 & 0.603 \\
Exclude OP and ZEC & 41 & 0.294 & 0.290 & 0.438 \\
Exclude shorter histories & 36 & 0.252 & 0.293 & 0.804 \\
\bottomrule\end{tabular}
\end{table}

The padded bootstrap mean is $0.279$, about $33.2\%$ above the point estimate. Its 2.5th percentile, $0.214$, is itself above the observed padded coefficient. This is why we do not interpret those percentile bounds as ordinary uncertainty limits for the population alignment.

\subsection{Document replacement and composition have distinct effects}
\label{sec:composition}
Let $L$ be the 34 assets shared by the old and screened document snapshots. Four legs distinguish changes in text from changes in rows: A uses the old 37-asset sample; B restricts its texts to $L$; C uses screened texts on the same $L$; D uses the full screened 43-asset sample. The market statistics and their scaling are held fixed within each volume specification. A$\to$B and C$\to$D change composition; B$\to$C changes only the eight replaced text rows. These contrasts isolate particular perturbations of this archive. They are not a randomised experiment in document contamination and do not estimate an average contamination effect across corpora.

\begin{table}[H]\centering
\caption{Four-leg decomposition, with $p$-values in parentheses.}\label{tab:composition}
\begin{tabular}{llrrr}\toprule
Volume & Leg & $n$ & Matched $\phi$ & Padded $\phi$\\\midrule
USD & A & 37 & 0.268 (0.777) & 0.200 (0.850) \\
USD & B & 34 & 0.299 (0.606) & 0.221 (0.620) \\
USD & C & 34 & 0.278 (0.755) & 0.207 (0.778) \\
USD & D & 43 & 0.280 (0.507) & 0.210 (0.546) \\
Base-token & A & 37 & 0.304 (0.528) & 0.246 (0.367) \\
Base-token & B & 34 & 0.345 (0.293) & 0.255 (0.306) \\
Base-token & C & 34 & 0.335 (0.371) & 0.246 (0.403) \\
Base-token & D & 43 & 0.303 (0.366) & 0.223 (0.434) \\
\bottomrule\end{tabular}
\end{table}

On USD turnover the padded statistic changes by $+0.0218$ when three assets are removed, $-0.0140$ when eight documents are replaced on the fixed 34-asset set, and $+0.0024$ when nine assets are added. The corresponding base-volume changes are $+0.0090$, $-0.0088$ and $-0.0229$. The two market specifications therefore imply different relative magnitudes and a different sign for the final composition step. A general claim that composition dominates contamination, or that the documents alone manufactured the entire ordering, is not licensed by these contrasts. All four legs under both specifications remain non-significant.

For an entity diagnostic define $\Delta_i=\phi_{\rm padded}(C,S)-\phi_{\rm padded}(C_{-i},S_{-i})$. The coefficient is refitted after each deletion; $\Delta_i$ measures influence on this fitted statistic, not whether token $i$ conforms to its whitepaper. On the full screened USD sample impacts range from $-0.0179$ to $+0.0120$. Their absolute size and ordering depend on sample and specification. We therefore report them continuously and do not divide tokens into ``helpers'' and ``hurters'' using a $\pm0.01$ threshold. A row-normalised short document is not an empty observation, and a threshold crossing does not identify a causal leverage mechanism.

\begin{figure}[H]\centering
\includegraphics[width=\textwidth]{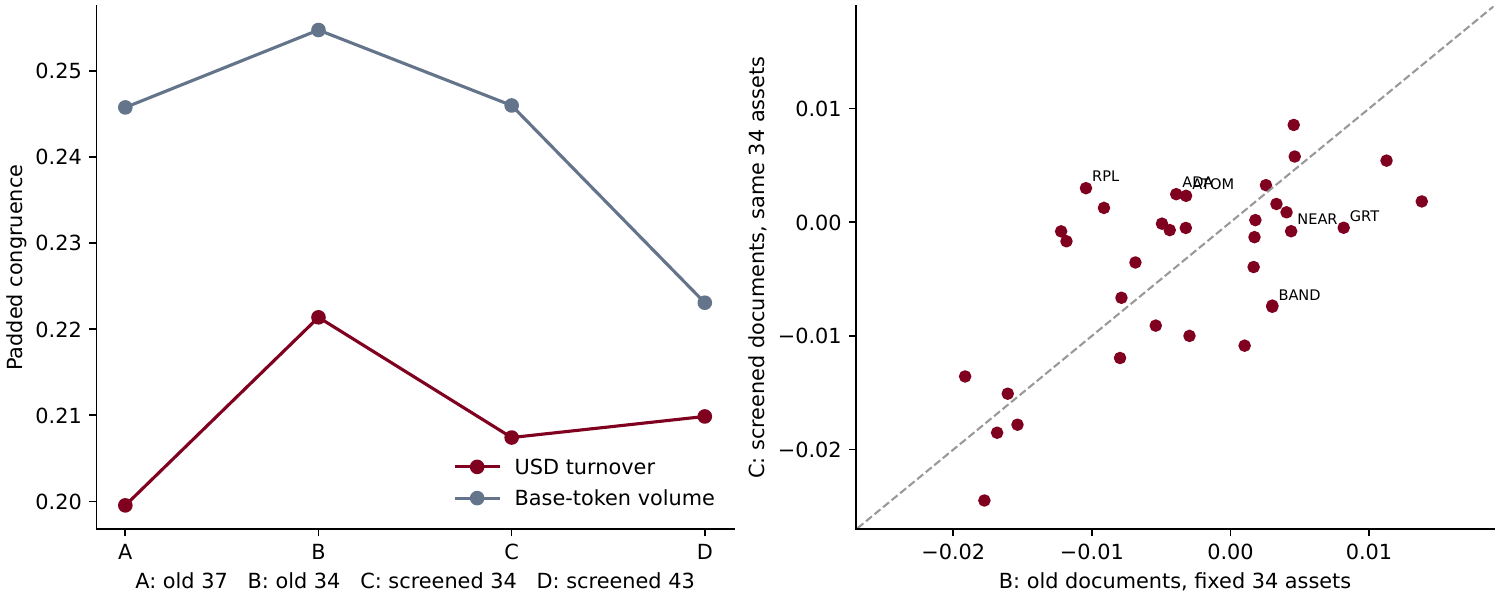}
\caption{Composition and document-content perturbations. Left: all four legs on each fixed volume specification. Right: paired leave-one-out impacts for B and C on the same 34 assets and the same USD market matrix. A point on the diagonal has unchanged influence after document replacement. Neither panel compares mismatched specifications as though the difference were a pure contamination effect.}\label{fig:composition}
\end{figure}

\section{Conditional Sensitivity of the Numerical Alignment Stage}
\label{sec:power}
The numerical simulations use the centred $43\times7$ primary USD-statistics matrix $S$ as a fixed carrier. For injected parameter $\theta>0$, each carrier column is corrupted by independent Gaussian noise with variance $\operatorname{var}(S_j)(1-\theta_{\rm eff}^2)/\theta_{\rm eff}^2$. For the ideal scenario $\theta_{\rm eff}=\theta$; for agreement scenario $r<1$, $\theta_{\rm eff}=\theta\sqrt{0.95r}$. The constant 0.95 is an assumed carrier-reliability factor retained for continuity, not an empirical estimate. At $\theta=0$ the simulated text is independent noise, avoiding a numerically large pseudo-signal.

We compare two noise geometries. In the subspace condition, a random orthonormal embedding maps the seven corrupted carrier columns into ten text dimensions, leaving no noise in the three-dimensional complement. In the full-space condition, independent Gaussian noise is added in those three complementary directions, with per-direction variance equal to the mean noise variance in the seven carrier directions. The latter is a sensitivity scenario, not an estimate of real classifier errors. Both conditions are reduced back to seven principal components and tested by the same fitted statistic.

The grid is $\theta\in\{0,.2,.3,.5,.65,.8\}$ and $r\in\{1,.50,.314,.07\}$, with 300 independent replications per cell, 199 permutations per replication, deterministic cell seeds starting at 20260912, and rejection at $p<.05$ using the add-one permutation convention. There are 48 cells and 14,400 synthetic datasets. The largest possible Monte Carlo standard error for a cell rejection rate is about 0.029. Table~\ref{tab:power} reports selected cells; all cells and seeds are in \texttt{power.json}.

\begin{table}[H]\centering
\caption{Conditional rejection rates on the USD carrier. S = subspace noise; F = full-space noise.}\label{tab:power}
\begin{tabular}{llrrrr}\toprule
Noise & $r$ & $\theta=0$ & $\theta=.3$ & $\theta=.5$ & $\theta=.8$\\\midrule
S & 1.000 & 0.053 & 0.893 & 1.000 & 1.000 \\
S & 0.500 & 0.050 & 0.447 & 0.973 & 1.000 \\
S & 0.314 & 0.040 & 0.233 & 0.803 & 1.000 \\
S & 0.070 & 0.057 & 0.073 & 0.177 & 0.420 \\
F & 1.000 & 0.040 & 0.783 & 1.000 & 1.000 \\
F & 0.500 & 0.023 & 0.317 & 0.900 & 1.000 \\
F & 0.314 & 0.053 & 0.250 & 0.690 & 1.000 \\
F & 0.070 & 0.057 & 0.077 & 0.147 & 0.343 \\
\bottomrule\end{tabular}
\end{table}

\begin{figure}[H]\centering
\includegraphics[width=\textwidth]{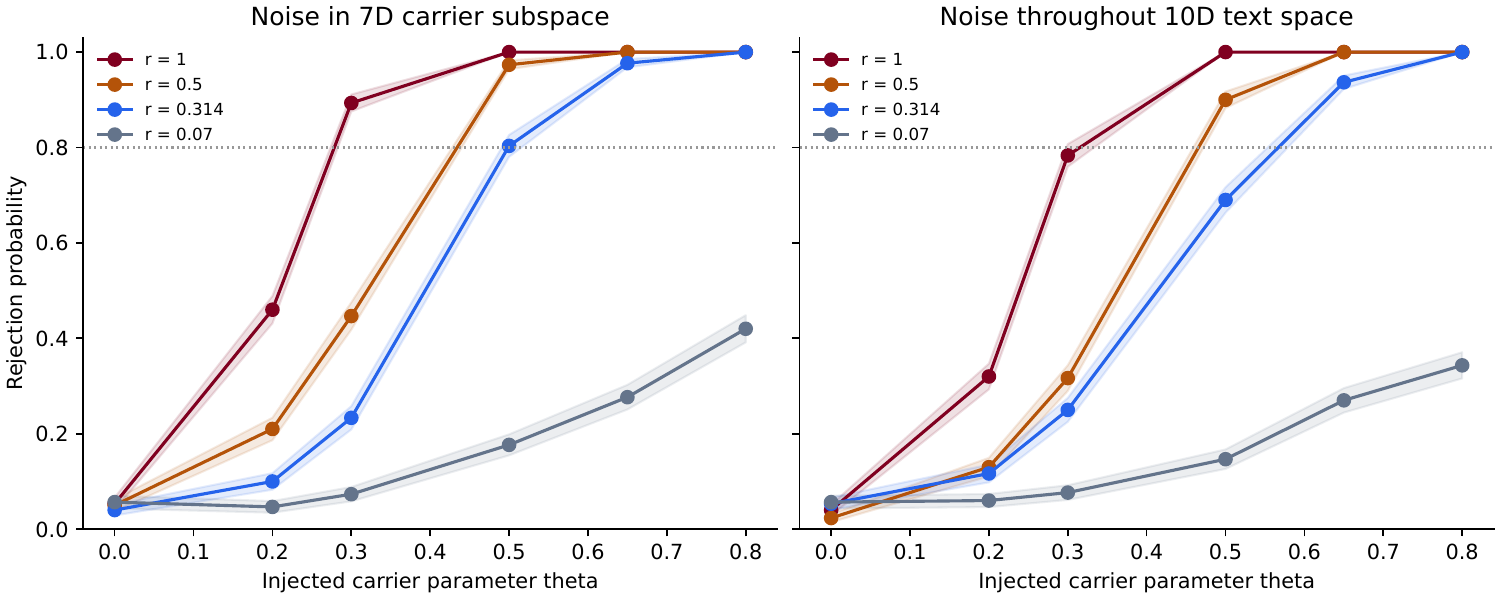}
\caption{Sensitivity of the numerical alignment procedure under two assumed noise geometries. Shading is plus/minus one Monte Carlo standard error. The horizontal axis is the injected carrier parameter $\theta$, not an empirical estimate of text validity or the fitted Procrustes statistic.}\label{fig:power}
\end{figure}

At $r=.07$ and $\theta=.8$, detection is $42.0\%$ with subspace noise and $34.3\%$ with full-space noise. At $r=.314$ and $\theta=.5$ it is $80.3\%$ versus $69.0\%$. Thus the noise structure can affect whether the same injected parameter appears well detected. These contrasts do not imply that the full-space model is true; they demonstrate dependence on assumptions omitted from an unconditional power claim.

The fitted coefficient and the injected parameter are different scales. Rotation fitting, dimensional reduction and finite-sample noise can produce an average fitted coefficient materially above the injected value, especially near zero. Comparing the observed $0.280$ directly with a simulated $\theta$ threshold would confuse a statistic with a data-generating parameter. We consequently do not report the interpolated thresholds of earlier drafts as exclusion bounds, do not say the study rules out $\phi\geq0.70$, and do not describe the estimator as recovering all injected effects exactly.

The values of $r$ are scenarios informed by disagreement between automated instruments. The frozen BART--embedding, BART--LLM and embedding--LLM pooled Pearson correlations are respectively $0.070$, $0.475$ and $0.395$; their average is $0.314$. Correlations within categories differ from pooled cross-cell correlations, and neither is ground-truth reliability. Shared bias can produce agreement and different operational definitions can produce disagreement. No simulation here passes a document through the classifier or validates its output against human labels. End-to-end sensitivity to economically meaningful claims remains unmeasured.

\section{Discussion}
\label{sec:discussion}
The finding is a non-detection under an explicit descriptive representation. It does not establish that narratives are irrelevant to cryptocurrency prices, that markets ignore fundamentals, or that functional commitments and market structure are independent. The chosen topic representation may omit assertions relevant to prices; the market representation omits many economically relevant quantities; and the two representations may be poorly measured even when a relation exists between the underlying constructs. The limited and selected sample further restricts generalisation.

The corpus episode makes a separate, narrower point. File identity, asset inclusion and market units must be tracked separately if differences between manuscript versions are to be interpreted. A failed download, a wrong-project document and the inclusion of three additional assets are different interventions on an analysis. In this archive the direction and size of some changes also depend on whether volume is measured in tokens or approximate dollars. The correction is not evidence for a universal ranking of these risks. It is evidence that the original causal attribution exceeded what the available contrasts identified.

Historical validity remains unresolved. Removing OP and ZEC does not detect alignment, but this does not date-validate every remaining text. A stronger historical design would freeze a versioned document corpus available before or during the market window and state inclusion rules before inspecting the outcomes. It would also use matched market periods or model differing coverage explicitly. Those improvements address design validity; a larger sample alone cannot repair mislabeled constructs or postperiod documents. No sample-size simulation in this release establishes that 50 or any other particular number of projects would suffice.

A more informative next measurement study would pair domain-informed human coding of assertions with model outputs, evaluate category-specific errors, retain unclosed chunk scores and track document versions. On the market side, token supply and independent market-capitalisation data would permit a genuine size control, while on-chain adoption and event-window outcomes would test more specific mechanisms. Such a design should state the target estimand before selecting an alignment statistic and calibrate its inferential procedure under plausible text and market error models. This paper establishes neither those mechanisms nor that expanded validation; it provides the inspectable baseline and records which interpretations the current data fail to support.

\section{Conclusions}
Using approximate USD turnover and the frozen 43-asset topic profiles, dimension-matched congruence is $0.280$ with permutation $p=0.507$ and null mean $0.284$. Exclusions for Bitcoin, known postperiod documents and shorter market histories do not produce a significant result. Replacing documents on a fixed asset set has a measurable but specification-dependent effect, and the earlier contamination-only entity narrative is withdrawn. Numerical simulations show sensitivity conditional on an assumed carrier, signal parameter and noise geometry; they validate neither historical corpus availability nor the text instrument. The empirical contribution is a reproducible non-detection and an explicit account of the measurement and inference limits attached to it.

\section*{Declarations}
\paragraph{Status.} Preprint. No acceptance by Financial Innovation is asserted.
\paragraph{Conflict of Interest and Funding.} The author declares no competing interests. This research received no external funding.
\paragraph{Data and Code.} Code, frozen score matrices and source data are associated with \url{https://github.com/dissensus-ai/whitepaper-claims}. The numerical results are archived at \href{https://github.com/dissensus-ai/whitepaper-claims/tree/679015c692f7678df4e5d240393b7f49ae1f49a6}{analysis commit \texttt{679015c}}. This revision's reproducible analysis is in \path{scripts/release_20260912/verify_release.py}; outputs are in \path{outputs/release_20260912/}. The accompanying source archive contains the manuscript and its used figures. The code/data release and provenance manifest are separate inspection artifacts. The repository's existing licence is CC BY 4.0; rights in third-party documents remain with their respective owners.
\paragraph{AI Assistance.} Earlier drafts and code received assistance from Claude (Anthropic); earlier adversarial checks used ChatGPT (OpenAI) and citation tools including scite. This revision used Codex (OpenAI) for source/code auditing, numerical reruns, implementation changes, scientific adjudication, manuscript revision and packaging. The author is responsible for the released work and its claims.
\paragraph{Author Contributions.} Murad Farzulla is the sole author.

\appendix
\section{Asset and Input Audit}
The aligned sample is: AAVE, ADA, ALGO, API3, APT, ARB, ATOM, AVAX, BAND, BTC, COMP, CRV, DOT, EGLD, ENJ, ETH, FIL, FTM, GRT, HBAR, ICP, IMX, LDO, LINK, LPT, MANA, MKR, NEAR, OCEAN, OP, POL, RENDER, RPL, SAND, SC, SNX, SOL, STORJ, SUI, TRB, UNI, XMR, ZEC. The six market assets without an analysed current text profile are AXS, ENS, GALA, LIT, SUSHI and YFI. The source archive preserves the study's numerical representation; it does not certify all source-document dates.

\begin{longtable}{lrrr}\caption{Archived market coverage and analysed text chunks. Chunk counts are extraction units, not independently coded claims.}\label{tab:inputs}\\
\toprule Asset & Hourly bars & Text chunks & Stored file (kB)\\\midrule\endfirsthead
\toprule Asset & Hourly bars & Text chunks & Stored file (kB)\\\midrule\endhead
AAVE & 17,543 & 30 & 1168 \\
ADA & 17,543 & 96 & 1768 \\
ALGO & 17,543 & 226 & 671 \\
API3 & 17,543 & 96 & 493 \\
APT & 17,543 & 72 & 474 \\
ARB & 15,584 & 73 & 405 \\
ATOM & 17,543 & 91 & 89 \\
AVAX & 17,543 & 46 & 255 \\
BAND & 17,543 & 37 & 36 \\
BTC & 17,543 & 23 & 184 \\
COMP & 17,543 & 21 & 189 \\
CRV & 17,543 & 10 & 274 \\
DOT & 17,543 & 145 & 4076 \\
EGLD & 17,543 & 96 & 814 \\
ENJ & 17,543 & 47 & 3457 \\
ETH & 17,543 & 95 & 941 \\
FIL & 17,543 & 98 & 629 \\
FTM & 17,543 & 84 & 827 \\
GRT & 17,543 & 23 & 202 \\
HBAR & 17,543 & 31 & 900 \\
ICP & 17,543 & 132 & 1212 \\
IMX & 17,543 & 48 & 1737 \\
LDO & 17,543 & 22 & 26324 \\
LINK & 17,543 & 94 & 735 \\
LPT & 17,543 & 67 & 62 \\
MANA & 17,543 & 28 & 4557 \\
MKR & 17,543 & 40 & 300 \\
NEAR & 17,543 & 134 & 1383 \\
OCEAN & 13,130 & 97 & 1515 \\
OP & 17,543 & 65 & 611 \\
POL & 2,630 & 49 & 427 \\
RENDER & 3,808 & 25 & 351 \\
RPL & 17,127 & 46 & 935 \\
SAND & 17,543 & 89 & 12646 \\
SC & 17,543 & 26 & 156 \\
SNX & 17,543 & 58 & 3813 \\
SOL & 17,543 & 50 & 689 \\
STORJ & 17,543 & 212 & 750 \\
SUI & 14,604 & 72 & 632 \\
TRB & 17,543 & 19 & 415 \\
UNI & 17,543 & 31 & 316 \\
XMR & 9,962 & 50 & 446 \\
ZEC & 17,543 & 701 & 3312 \\
\bottomrule\end{longtable}

\section{Classifier Labels and Agreement}
The ten labels passed to the classifier are the complete sentences in \texttt{src/nlp/taxonomy.py}. Their labels describe store of value, payment systems, smart contracts, decentralised finance, governance, scalability, privacy, interoperability, storage and oracles. They do not distinguish endorsement from criticism, fulfilled commitments from aspirations, or the project's own function from a cited comparator. The normalised matrix therefore records relative topic emphasis.

\begin{figure}[H]\centering
\includegraphics[width=.96\textwidth]{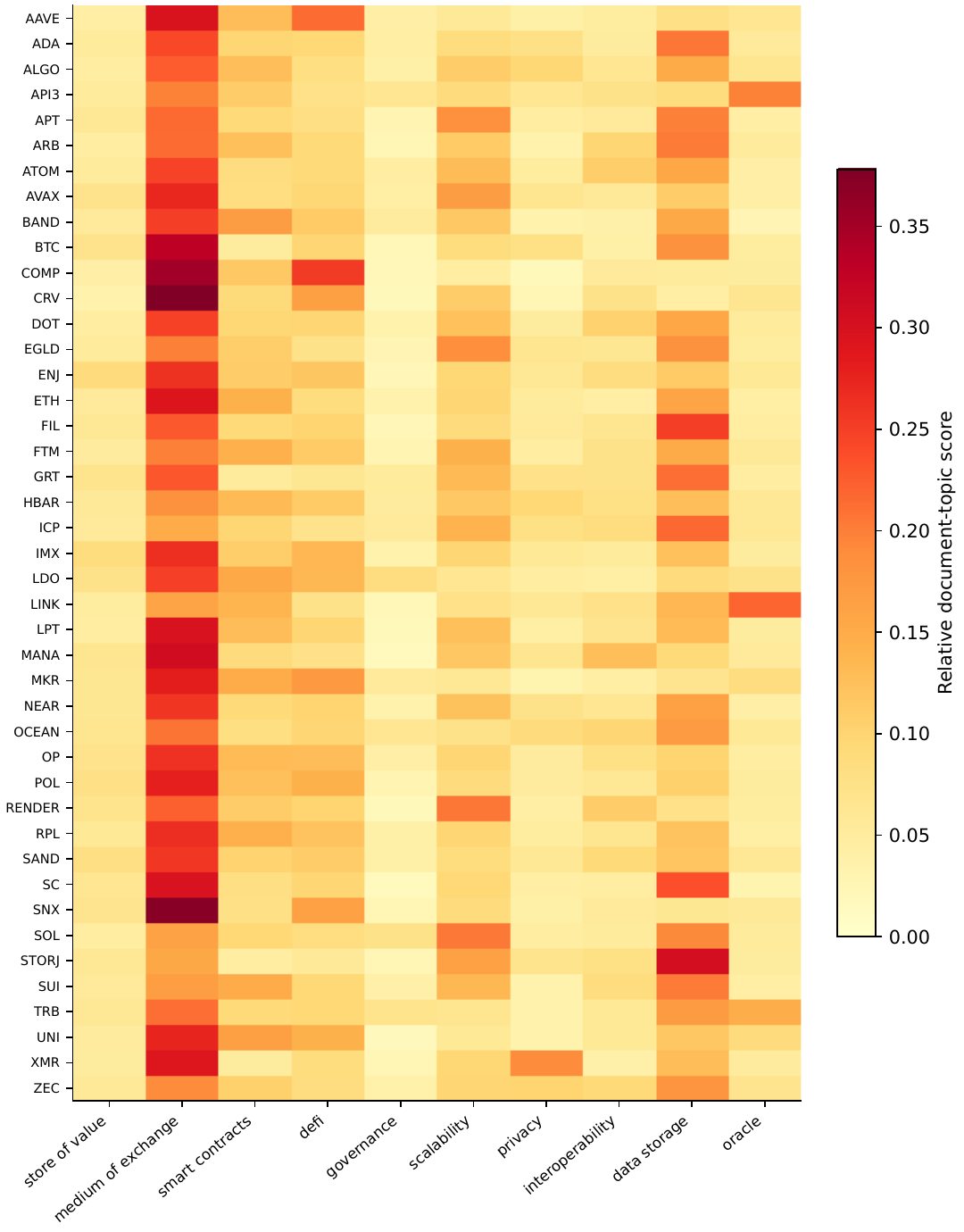}
\caption{Frozen topic-emphasis matrix used in the release analysis. Rows sum to one. Darker cells represent relative classifier scores, not probabilities of realised economic functions.}\label{fig:topics}
\end{figure}

The mean category-specific pairwise correlations range from $0.208$ for smart contracts to $0.815$ for DeFi. Most assets receive medium of exchange as their highest-scoring category (36 of 43); four are storage-modal, two oracle-modal and one scalability-modal. This concentration constrains the representation's ability to discriminate between projects even before market alignment is considered. It motivates validation against independently coded financial constructs rather than interpreting inter-model agreement as a sufficient reliability estimate.

\section{Reproduction and Version Boundaries}
Run \path{python scripts/release_20260912/verify_release.py --power} from a copy of the code release. The script rebuilds the market matrices from Parquet bars, checks the matched coefficient with an independent vectorised implementation, computes permutation distributions, reruns the four-leg comparisons under two volume specifications, performs sample exclusions and the padded bootstrap, writes a document/chunk manifest, and runs the 48 simulation cells. It uses a fixed seed for each stochastic component and stores exact exceedance counts. The environment lock and checksums accompanying the release record the versions used for this revision.

The analysis begins with frozen classifier profiles. Re-executing the NLP stage requires model weights and the full original model stack, and was not part of the September numerical verification. The legacy \texttt{run\_full\_pipeline.py} entry point retains the earlier base-token volume workflow and should not be used as though it generated this revision's primary tables. Legacy figures and outputs remain available as historical artifacts, but claims resting on unverified combinations of estimator, sample or specification have been removed from this manuscript. The journal-held August manuscript has not been silently overwritten; this is a separately dated correction candidate.

\begin{thebibliography}{99}
\bibitem[Lewis et~al.(2020)Lewis, Liu, Goyal, Ghazvininejad, Mohamed, Levy,
  Stoyanov, and Zettlemoyer]{lewis2020bart}
Mike Lewis, Yinhan Liu, Naman Goyal, Marjan Ghazvininejad, Abdelrahman Mohamed,
  Omer Levy, Veselin Stoyanov, and Luke Zettlemoyer.
\newblock {BART}: Denoising sequence-to-sequence pre-training for natural
  language generation, translation, and comprehension.
\newblock In \emph{Proceedings of the 58th Annual Meeting of the Association
  for Computational Linguistics}, pages 7871--7880, 2020.
\newblock \doi{10.18653/v1/2020.acl-main.703}.

\bibitem[Liu and Tsyvinski(2021)]{liu2021risks}
Yukun Liu and Aleh Tsyvinski.
\newblock Risks and returns of cryptocurrency.
\newblock \emph{The Review of Financial Studies}, 34\penalty0 (6):\penalty0
  2689--2727, 2021.
\newblock \doi{10.1093/rfs/hhaa113}.

\bibitem[Liu et~al.(2022)Liu, Tsyvinski, and Wu]{liu2022common}
Yukun Liu, Aleh Tsyvinski, and Xi~Wu.
\newblock Common risk factors in cryptocurrency.
\newblock \emph{The Journal of Finance}, 77\penalty0 (2):\penalty0 1133--1177,
  2022.
\newblock \doi{10.1111/jofi.13119}.

\bibitem[Lorenzo-Seva and ten Berge(2006)]{lorenzo2006tuckers}
Urbano Lorenzo-Seva and Jos~MF ten Berge.
\newblock Tucker's congruence coefficient as a meaningful index of factor
  similarity.
\newblock \emph{Methodology}, 2\penalty0 (2):\penalty0 57--64, 2006.
\newblock \doi{10.1027/1614-2241.2.2.57}.

\bibitem[Sch{\"o}nemann(1966)]{schonemann1966generalized}
Peter~H Sch{\"o}nemann.
\newblock A generalized solution of the orthogonal {P}rocrustes problem.
\newblock \emph{Psychometrika}, 31\penalty0 (1):\penalty0 1--10, 1966.
\newblock \doi{10.1007/BF02289451}.

\bibitem[Shiller(2017)]{shiller2017narrative}
Robert~J Shiller.
\newblock Narrative economics.
\newblock \emph{American Economic Review}, 107\penalty0 (4):\penalty0
  967--1004, 2017.
\newblock \doi{10.1257/aer.107.4.967}.

\bibitem[Williams et~al.(2018)Williams, Nangia, and Bowman]{williams2018broad}
Adina Williams, Nikita Nangia, and Samuel Bowman.
\newblock A broad-coverage challenge corpus for sentence understanding through
  inference.
\newblock In \emph{Proceedings of the 2018 Conference of the North American
  Chapter of the Association for Computational Linguistics: Human Language
  Technologies, Volume 1 (Long Papers)}, pages 1112--1122. Association for
  Computational Linguistics, 2018.
\newblock \doi{10.18653/v1/N18-1101}.

\bibitem[Yin et~al.(2019)Yin, Hay, and Roth]{yin2019benchmarking}
Wenpeng Yin, Jamaal Hay, and Dan Roth.
\newblock Benchmarking zero-shot text classification: Datasets, evaluation and
  entailment approach.
\newblock In \emph{Proceedings of the 2019 Conference on Empirical Methods in
  Natural Language Processing and the 9th International Joint Conference on
  Natural Language Processing (EMNLP-IJCNLP)}, pages 3914--3923. Association
  for Computational Linguistics, 2019.
\newblock \doi{10.18653/v1/D19-1404}.
\end{thebibliography}
\end{document}